# In-plane Selective Area InSb-Al Nanowire Quantum Networks

Roy L.M. Op het Veld[1,2†], Di Xu[2†], Vanessa Schaller[2], Marcel A. Verheijen[1,3], Stan M.E. Peters[1], Jason Jung[1], Chuyao Tong[1], Qingzhen Wang[2], Michiel W.A. de Moor[2], Bart Hesselmann[2], Kiefer Vermeulen[2], Jouri D.S. Bommer[2], Joon Sue Lee[4], Andrey Sarikov[5,6], Mihir Pendharkar[9], Anna Marzegalli[7] Sebastian Koelling[1], Leo P. Kouwenhoven[2,8], Leo Miglio[5], Chris J. Palmstrom[4,9,10], Hao Zhang[11,12,2\*], Erik P.A.M. Bakkers[1\*]

[1] Department of Applied Physics, Eindhoven University of Technology, 5600MB Eindhoven, The Netherlands
[2] QuTech and Kavli Institute of Nanoscience, Delft University of Technology, 2600GA Delft, The Netherlands
[3] Eurofins Materials Science Eindhoven, High Tech Campus 11, 5656AE Eindhoven, The Netherlands
[4] California NanoSystems Institute, University of California, Santa Barbara, California 93106, USA.
[5] L-NESS and Dept. of Materials Science, University of Milano-Bicocca, I-20125 Milano, Italy.
[6] V. Lashkarev Institute of Semiconductor Physics, National Academy of Sciences of Ukraine, Kiev, Ukraine.
[7] L-NESS and Dept. of Physics, Politecnico di Milano, I-22100 Como, Italy
[8] Microsoft Quantum Lab Delft, 2600GA Delft, The Netherlands
[9] Electrical and Computer Engineering, University of California, Santa Barbara, California 93106, USA
[10] Materials Department, University of California, Santa Barbara, California 93106, USA
[11] State Key Laboratory of Low Dimensional Quantum Physics, Department of Physics, Tsinghua University, Beijing 100084, China
[12] Beijing Academy of Quantum Information Sciences, Beijing 100193, China

† These authors contributed equally to this work.

\* Corresponding author e-mail: hzquantum@mail.tsinghua.edu.cn, e.p.a.m.bakkers@tue.nl

**Strong spin-orbit semiconductor nanowires coupled to a superconductor are predicted to host Majorana zero modes. Exchange (braiding) operations of Majorana modes form the logical gates of a topological quantum computer and require a network of nanowires. Here, we develop an in-plane selective-area growth technique for InSb-Al semiconductor-superconductor nanowire networks with excellent quantum transport properties. Defect-free transport channels in InSb nanowire networks are realized on insulating, but heavily mismatched InP substrates by 1) full relaxation of the lattice mismatch at the nanowire/substrate interface on a (111)B substrate orientation, 2) nucleation of a complete network from a single nucleation site, which is accomplished by**

**optimizing the surface diffusion length of the adatoms. Essential quantum transport phenomena for topological quantum computing are demonstrated in these structures including phase-coherent transport up to 10 μm and a hard superconducting gap accompanied by 2*e*-periodic Coulomb oscillations with an Al-based Cooper pair island integrated in the nanowire network.**

Indium-Antimonide (InSb) and Indium-Arsenide (InAs) nanowires are promising candidates for realizing Majorana-based topological quantum computers.[1] InSb, in particular, is interesting for its high electron mobility, strong spin-orbit coupling and large Landé g-factor.[2-5] Over the past years, extensive efforts have been made to improve the quality of InSb nanowires, grown by the vapor-liquid-solid (VLS) mechanism.[6, 7] A major achievement is the clean and epitaxial semiconductor-superconductor interface relying on *in-situ* metal evaporation and complex substrate processing.[8, 9] This has been critical for showing the first quantized conductance of Majorana zero-modes, a milestone towards braiding experiments.[10] To braid these Majorana states, scalable nanowire networks with a high degree of interconnectivity are required.[11] Recently, out-of-plane growth of InSb nanowire networks has been demonstrated by merging multiple wires during VLS growth. Phase-coherent and ballistic transport have been observed, demonstrating the high quality of these network structures.[8, 12] This technique, however, requires predefined positioning of the nanowires with nanometre accuracy in order to form networks, leading to a drastic decrease in yield with increasing network complexity. Moreover, merging of VLS nanowires inevitably forms a widening of the nanowire diameter at and around the junction with a 75% chance of a defect forming at the junction, which negatively affects the one-dimensionality of the system.[8]

Here, we demonstrate a more scalable approach, using an in-plane selective-area growth (SAG) technique (i.e. parallel to the substrate surface), that relies on a template or mask to selectively grow one semiconductor material on top of another.[13-19] Our technique has several advantages over out-of-plane growth. First, the flexibility of network designs is significantly enhanced, as the preferred design can be written and etched directly into a mask enabling complex structures. Second, the growth can be confined within the mask, keeping the entire structure one-dimensional with an easily controllable and constant cross-section size. Finally, the technique excels in scalability, readily allowing for the growth of complex structures suggested for Majorana braiding experiments in a variety of theoretical proposals.[11, 20, 21] The large lattice mismatch between InSb and any other III-V semiconductor substrate material,[22] however, poses an important challenge making it difficult to grow defect-free InSb nanowires on large-bandgap or insulating substrates. Furthermore, the disorder created by the lattice mismatch can be detrimental to the topological protection of Majorana states.[23] Most of the previous SAG studies have focused on an InAs-based material system for nanowire networks [14-17, 19] which has a smaller lattice mismatch with InP or GaAs substrates. InSb nanowires have been grown by SAG on (100) GaAs substrates but the electron transport demonstrated is diffusive.[18] Here, we demonstrate an in-plane SAG technique for scalable and high quality InSb nanowire networks, which shows all the relevant quantum transport properties (e.g. long coherence length and excellent induced superconducting properties) necessary for topological qubits.

In this study we use InP as a substrate because it has a type I band alignment with InSb and becomes semi-insulating at low temperatures. The large band gap of 1.34 eV compared to 0.17 eV of InSb ensures good confinement.[24] The lattice mismatch between these materials is 10.4%, making large scale defect-free growth of InSb a challenge.[25] It is important to have defect-free single crystalline material to obtain a high carrier mobility by reducing electron scattering.[26] On a (100) substrate, many stacking faults are expected to form inclined to the substrate (along the {111} planes) to relax the lattice strain.[13] To avoid the formation of these defects, we use InP (111)B as growth substrate. We will show that strain from the lattice mismatch is relieved directly at the substrate/nanowire interface. Moreover, atomically flat twin planes can form parallel to the substrate, after which the nanowires grow completely defect-free, separating the mismatch-induced disorder from the nanowire top part. On a (111)B substrate, growth nuclei with an odd and even number of horizontal twins will have a 180º-rotated crystallographic orientation with respect to each other. When these nuclei merge by lateral growth, a defect is formed at the interface; details of the atomic structure of such a defect can be found in the supplementary figure S1. These defects may act as scattering sites for electrons.[26] Therefore, it is important to enable growth of a complete, in-plane network structure from a single nucleation site. For this, a large surface diffusion length of the precursor material is required as well as a low nucleation probability. Here, we show that metalorganic vapour-phase epitaxy (MOVPE) grown InSb in-plane selective area networks (InSANe) can indeed generate complicated 1D networks from a single nucleation site.

First, the desired structures are etched into a 20-nm-thick $Si_xN_y$ mask (see figure 1b and S2), aligned to four different crystal directions, the [-211], [-101], [-1-12] and [0-11], on an InP (111)B substrate (see figure 1a). These four orientations are all three-fold symmetric, such that networks can be grown with angles of 30, 60 and 90 degrees between two nanowires (see figure S3). The growth starts by forming a nucleus in one of the lines, which develops into an InSb island over time (figure S4). All nuclei are terminated with a {111}B top facet and {110} side facets (figure 1b and 1d), regardless of the line orientation, implying that surface growth kinetics and lateral growth rates are identical for all <110> and <112> growth directions. For longer growth times, the growth continues in the lateral direction following the mask opening by growing {110} facets, as evidenced by atomic force microscopy (AFM) and transmission electron microscopy (TEM) (see supplementary information). When the structure is fully grown in the in-plane direction, the growth continues in the vertical <111>B direction. The height of the InSb network can be precisely tuned by the growth time (see figure S5). When the InSb grows higher than the mask, it also starts to expand in the lateral direction, especially at acute corners of the structure (figure S6), which is not ideal for transport measurements as the one-dimensional confinement is lost. An example of a network is shown in figure 1c (and figure S7), whose structure corresponds to the proposed geometry of a four-topological-qubit device.[20] The {110} planes of the original growth fronts are visible on the convex corners of this structure (figure 1d). These facets do not form on the concave corners due to the connection with another branch of the network (figure S6). Our platform provides freedom of design and scalability for a plethora of device structures (figure S8).

In order to minimize the formation of inclined defects, the number of nucleation sites in the mask openings per unit length (*n*) is investigated as a function of the input V/III ratio of the Tri-Methyl-Indium (TMIn) and Tri-Methyl-Antimony (TMSb) precursors. For this purpose, InSb is grown for a short time (1 minute) to observe the early stages of nucleation and understand the nucleation probability as a function of V/III ratio. *n* is determined for growth in the <112> and <110> oriented trenches for different V/III ratios (figure 2). The results show a clear decrease of *n*, and thus an increase of the diffusion length of the adatoms on the InP surface, with increasing V/III ratio. At the highest TMSb pressures, the nucleation of InSb islands is completely inhibited. The inset in figure 2a is a logarithmic plot of the same data up to a V/III ratio of 20,000, showing a decrease of *n* with increasing V/III ratio. We note that there is no parasitic growth on the mask under any of these conditions. The red-circle and black-square (blue and green triangle) data points are all taken with the same TMSb (TMIn) partial pressure and varying TMIn (TMSb) partial pressure respectively. Figure 2a shows that these datapoints all follow the same trend, indicating that not the total flow but the V/III ratio is important in the studied range. Figure 2b shows a scanning electron microscopy (SEM) image of nuclei grown using a low V/III ratio. The InSb islands are only tens of nanometres long and approximately 20 nm thick. Figure 2c shows an SEM image of a representative nucleus grown with a very high V/III ratio. Here, the InSb island is much longer (300 nm) and less than 20 nm high. By integrating over the total volume of all nuclei in a given structure, we find that the lateral growth rate is determined by the TMIn flux (with higher flux giving faster growth rates) and the V/III ratio (with higher V/III ratio giving slower growth rates). From these results, we conclude that Sb changes the surface energy on the InP substrate and enhances the surface diffusion of the In precursor material.[27] For the growth of large networks, a high V/III ratio is thus beneficial to have a minimum number of nucleation events. The largest single crystalline networks we have fabricated with our method have a wire diameter of around 60 nm with lengths of up to 11 µm (see figure S8d).

The structural quality of the in-plane wires has been investigated by transmission electron microscopy (TEM). Focused Ion Beam (FIB) sample preparation was used to cut out cross-sections parallel and perpendicular to the long axis of in-plane [-211] and [0-11] oriented nanowires. Figure 3a shows a cross-sectional view of a nanowire grown along the [-211] direction. The Energy-Dispersive X-Ray spectroscopy (EDX) elemental mapping (figure 3b) shows the InP substrate (green) with the InSb nanowire (red) grown slightly higher than the $Si_xN_y$ mask (blue, top indicated by a white dash on the right). A Fast Fourier Transform (FFT) of the InSb, the InSb/InP interface and the InP regions of the high resolution scanning TEM (STEM) image (figure 3d in green, blue and purple box respectively) show that the InSb and the InP are both single crystalline while the FFT pattern of the interface region displays a combination of two single crystalline patterns with a different lattice constant indicating that the materials are fully relaxed. An Inverse Fast Fourier Transform (IFFT) of a zoom-in on the InSb/InP interface (figure 3c) shows misfit dislocations and their confinement to the interface between the two materials. This real space image was constructed by filtering for the (110) periodicity in the FFT pattern of a HRSTEM image and subsequently creating an Inverse FFT image. In this image, the ratio of the number of InP/InSb atomic columns is 96/87 = 1.1034. The 10.34% decrease in vertical lattice planes is in good agreement with the reported value of a 10.4% lattice mismatch between InSb and InP [4], indicating that the heterostructure is fully relaxed (inelastically) at the interface in the lateral direction. Here, it should be mentioned that when imaging orthogonally to this crystal direction the InP/InSb interface cannot be imaged accurately due to the slight recess of the wire into the substrate. Most likely, along the long axis, misfit dislocation planes will also be present with the dislocation lines located at the interface, as we do not see vertically extending defects orthogonal to the long axis of the wires in TEM studies. To visualize the presence of horizontal twin planes in the InSb structures, we investigate a cross-section of a nanowire grown along the <110> direction (figure 3e). A zoom-in on the bottom part (figure 3f) reveals a set of two twin boundaries a few nanometres above the InP/InSb interface (figure 3g). Horizontal twin planes have in total been observed in 12 out of 17 TEM samples (6 in perpendicular cuts of <110> grown wires and 6 in parallel cuts of <112> grown wires, the other 5 TEM samples did not show horizontal twin planes). The twin planes are always located within a few nanometres above the InSb/InP interface and a single crystalline InSb nanowire always forms above the twin. Once a twin plane is formed in a nucleus it will be extended into the rest of the network by lateral growth. The complete relaxation of lattice strain at the nanowire/substrate interface (bottom part) followed by horizontal twin planes helps to separate the electron wavefunctions in the active region of the device from the interface disorder.[28] This effect allows us to fabricate in-plane InSb nanowires on InP that have quantum transport properties comparable to free-standing structures as demonstrated by the high quality quantum transport in the next section.

We now turn to the quantum transport properties of our InSb InSANe system to demonstrate its feasibility for topological quantum information processing. The key ingredient in the measurement-based gate operation and topological qubit readout is the phase coherence, which can be revealed by the Aharonov-Bohm (AB) effect.[11, 20] An SEM image of a fabricated InSANe device is shown in the inset of figure 4b. The magneto-conductance in figure 4a reveals the AB oscillations. The oscillation amplitude is ~20% of $e^2/h$ at 20 mK., about an order of magnitude larger than for previously reported InSb VLS nanowire network structures.[8] Therefore, higher harmonics (up to the third) in the Fast Fourier Transform (FFT) are observed (figure 4b). This means that the electron's phase coherence remains measurable after circulating through the loop 1.5 times, giving an estimated phase coherence length of more than 9 microns in this device. A second device, shown in the inset of figure 4d, shows an even larger AB amplitude (~60% of $e^2/h$ in figure 4c) and up to 5 harmonics in the FFT spectrum (figure 4d). As the circumference of the AB loop in this device is 2.7 µm, confirming the reproducibly high quality of our networks. The measured AB period matches the loop area in all measured devices, *i.e.* their periods equal h/ne (for the $n^{th}$ harmonic). We point out that this is the first time to observe higher order AB harmonics in nanowire loop structures [8, 14, 15, 19, 29], suggesting high quality material. This result corroborates that the lattice-mismatch-induced disorder at the nanowire-substrate interface has negligible effect on the phase coherent transport. Finally, we observe a sharp weak anti-localization (WAL) peak in the magneto-conductance of this device around zero magnetic field (figure 4e), indicating the strong spin-orbit nature of the InSb nanowire. Fitting this WAL curve requires a new theoretical model applicable for nanowire networks, which will be developed in future studies.

The next important step is to introduce superconductivity in the InSb InSANe system. In order to form superconducting contacts to create the semiconductor-superconductor hybrid networks needed for the formation of Majoranas, the InSANe InSb samples, after growth, are transferred from an MOVPE to an MBE system. Here, the surface oxides are removed using atomic-hydrogen cleaning under (ultra-high vacuum) UHV conditions followed by 7 nm aluminium deposition at a sample temperature of around 120K, leading to a clean and smooth InSb-Al interface (see figure S9).[8]

Since *in-situ* shadowing methods to selectively grow superconductors on these InSb in-plane structures are not yet developed, we exploit a reliable selective etching recipe to selectively etch Al on InSb. This novel fabrication recipe enables us to define the positions of tunnel barriers and the Al film by lithography, facilitating flexible device designs. Figure 5a shows such a device where part of the Al is selectively etched away, and a tunnel gate electrode is added to deplete the InSb wire locally. A super-gate is deposited on the superconducting region of the nanowire whose cross-section is shown in figure 5b. The differential conductance on this normal-nanowire-superconductor (N-NW-S) device reflects the quasi-particle density-of-states in the proximitized nanowire segment, *i.e.* the induced superconducting gap as shown in figure 5c with a line-cut in figure 5d. The sub-gap conductance reaches zero, indicating a hard gap, a necessary condition for topological protection. Magnetic field dependence of accidental quantum dot levels reveals an effective *g*-factor of 18.6 (figure S11), smaller than the bare InSb g-factor of 50 but significantly larger than that of Al ($|g|=2$), indicating the wavefunction hybridizes between InSb and Al.[28, 30] The measured hard gap, together with the effective *g*-factor defined by the coupling between Al and InSb, suggest that the electron wavefunction is mainly distributed near the top (close to Al) where the wire is single crystalline with no noticeable disorder,[31-33] suggesting that disorder at the InP/InSb interface has a negligible effect here.

Finally, we explore the transport of a hybrid InSb/Al island device (figure 5e), with a finite charging energy. The charging energy of the hybrid island can mediate the coupling between the two Majorana states for qubit operations and readout.[11, 20, 34] If the charging energy is less than the superconducting gap, the system ground state energetically 'favours' even number of electrons, *i.e.* all electrons on the island form Cooper pairs in the superconducting condensate.[35] In transport, each Coulomb blockade diamond then corresponds to 2 electrons (1 Cooper pair), as shown in figure 5f. The 2*e*-periodic Coulomb oscillation at zero bias (black curve) indicates negligible quasi-particle poisoning. A higher bias voltage can excite quasi-particles, resulting in the regular 1*e*-periodic Coulomb oscillations (red curve). Applying a magnetic field along the nanowire splits the 2*e*-peaks into 1*e*-peaks (figure 5g), with oscillating even/odd peak spacing (figure 5h). This 2*e* to 1*e* transition might be interpreted as the appearance of two Majorana states, which allows the coherent 'teleportation' of a single electron.[35] The oscillating peak spacing could be attributed to overlapping Majorana wavefunctions.[36] We note that a trivial explanation for the 2e to 1e transition based on Andreev bound states cannot be ruled out at this point.[37-39]

We have studied the growth dynamics of in-plane InSb nanowires on InP (111)B substrates. Despite the large mismatch between the wires and the substrate, defect-free single crystalline network channels are formed due to immediate strain-relaxation at the nanowire-substrate interface. Correspondingly, these in-plane InSb based devices show high quality quantum transport, with long phase coherence length, a hard superconducting gap, 2*e*-Coulomb blockade peaks and possible Majorana/Andreev signatures. The next step is to establish Majorana zero modes in these structures by performing key experiments like correlation and Majorana braiding. [40]

**Acknowledgments**

This work has been supported by the European Research Council (ERC HELENA 617256 and Synergy), the Dutch Organization for Scientific Research (NWO), and Microsoft Corporation Station-Q. We acknowledge Solliance, a solar energy R&D initiative of ECN, TNO, Holst, TU/e, imec and Forschungszentrum Jülich, and the Dutch province of Noord-Brabant for funding the TEM facility. The work at University of California, Santa Barbara was supported in part by Microsoft Research. We also acknowledge the Department of Energy (DE-SC0019274) and the use of facilities within the National Science Foundation Materials Research and Science and Engineering Center (DMR 11–21053) at the University of California, Santa Barbara. We thank Ghada Badawy and Ksenia Korzun for their careful reading of the manuscript.

**Author contributions**

R.L.M.O.H.V., S.M.E.P., J.J., C.T. and E.P.A.M.B. carried out the substrate processing and the InSb synthesis. D.X., V.S. Q.W., M.W.A.d.M., B.H., K.V., J.D.S.B., L.P.K. and H.Z., fabricated the devices and performed the transport measurements. J.S.L., M.P. and C.J.P. have deposited the superconductor. S.K. and J.J. have performed the FIB cuts for the TEM lamella and M.A.V. is responsible for the TEM analysis. A.S., A.M., and L.M. performed the theoretical modelling of defect formations. R.L.M.O.H.V., D.X., M.A.V., H.Z. and E.P.A.M.B. have authored the paper, with contributions from all authors. The authors declare no competing financial interests.

**Data availability**

The raw data and the data analysis codes that support the findings of this research are available at https://doi.org/10.5281/zenodo.4589484.

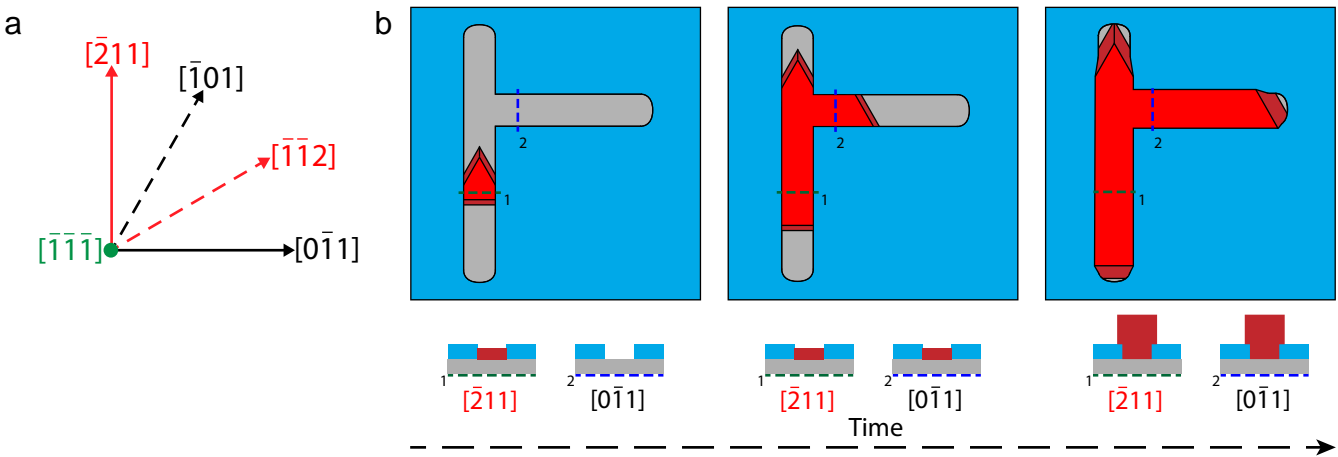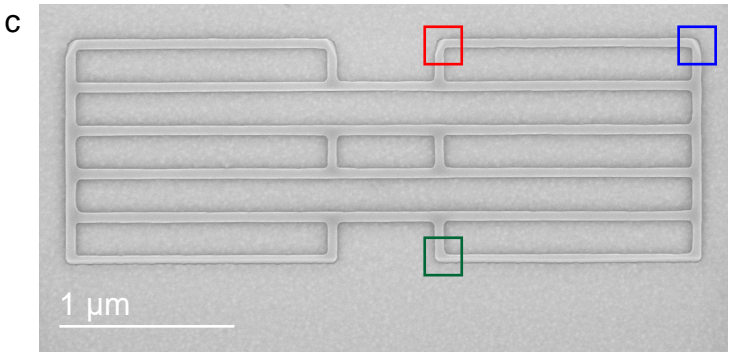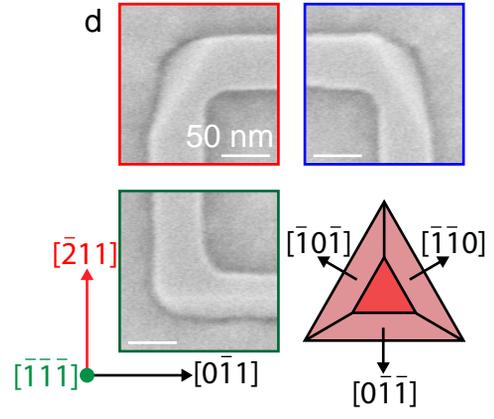

**Figure 1| Controlled growth of in-plane selective area InSb networks**.

**a:** the four growth directions on a (111)B substrate suitable for in-plane nanowire growth. **b:** Time evolution for growth of an in-plane InSb (red) nanowire network on an InP (111)B substrate (grey) with a 20 nm thick $Si_xN_y$ mask (light blue) for selectivity. The growth fronts remain the same during the growth of the network. Schematics of the cross-sections below show the relative height of the InSb compared to the height of the mask. The network grows much faster in the in-plane direction than the out-of-plane direction (see also figure S5). Cross-sections for both crystal directions are shown in figure 3. **c:** The in-plane InSb nanowires controllably grow in the <112> and the <110> families of crystal directions. **d:** Zoom-in on the corners of the structure in (c). The {110} facets that form the growth fronts are still visible at these edges.

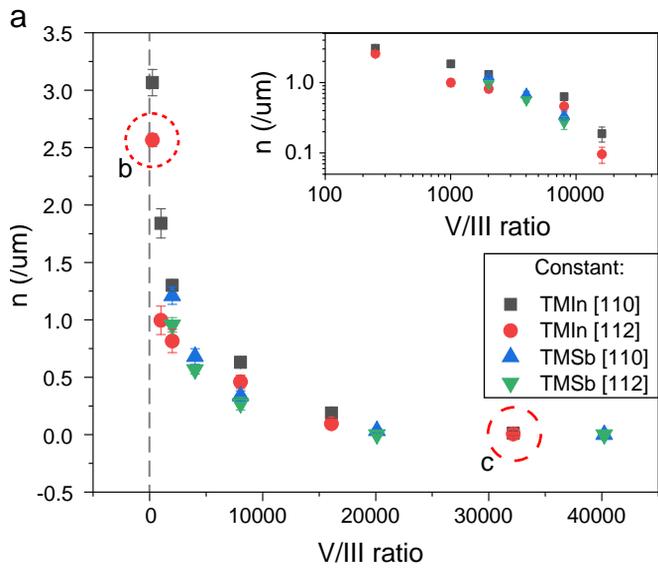
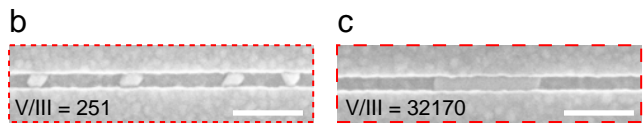

**Figure 2| Effect of V/III ratio on nucleation of InSb on InP.**

**a**: The average number of nucleation points *n* per micrometre trench length as a function of the V/III ratio during the growth for different crystal directions of the mask openings ([110] and [112]). The datapoint averages are gathered by analysing 500 µm length of mask opening (10 lines, each 50 µm long). Varying the TMSb (TMIn) flow with constant TMIn (TMSb) flow leads to the same value for *n* as a function of V/III, showing that the total flow does not influence the nucleation of the InSb on the InP. At a ratio of 40,000, the nucleation of InSb is completely inhibited by Sb adatoms on the InP surface. **Inset**: Logarithmic plot of the data up to a V/III ratio of 20,000. **b, c**: Top-view SEM image showing the different morphologies of, and distance between, InSb islands for low (b) and high (c) V/III ratios. Scalebar in **b, c** is 200 nm.

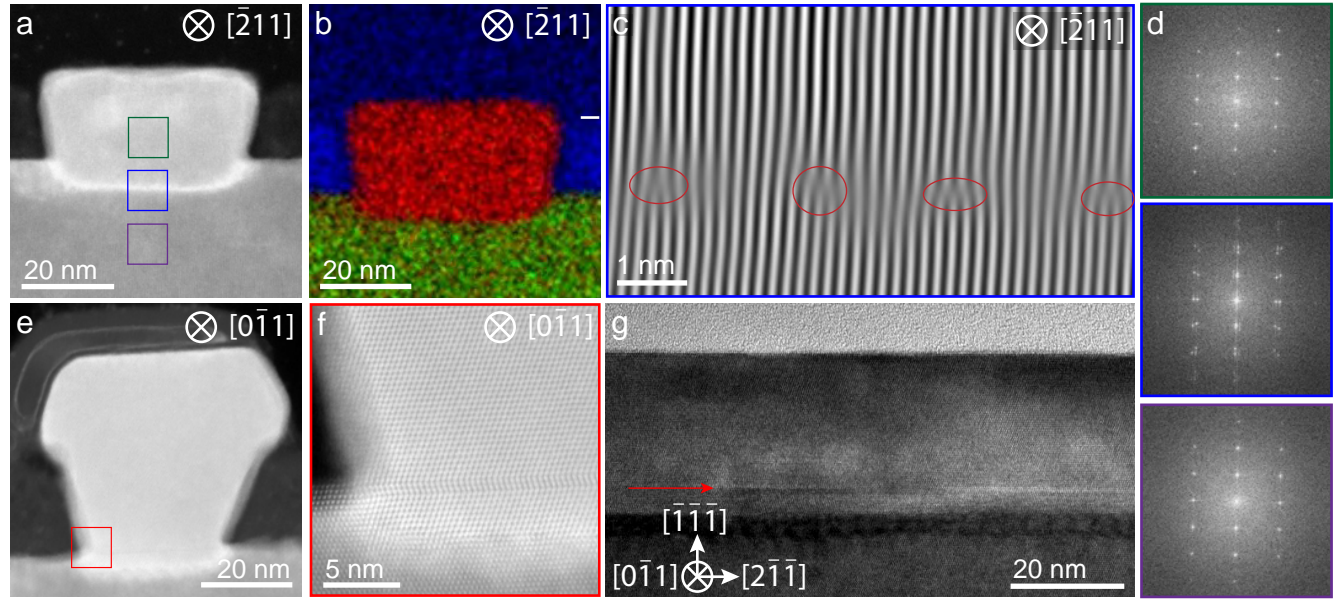

**Figure 3| Crystal quality analysis of InSb nanowires and the InSb/InP interface.**

**a:** Cross-section transmission electron microscope (TEM) image of an In-plane Selective Area (InSANe) InSb nanowire grown and imaged along the [-211] direction. **b:** EDX map of (a) with P (green), Sb (red) and Si (blue) depicted showing the InSb nanowire, the InP substrate and the $Si_xN_y$ mask. The top of the mask is indicated by a white dash on the right **c:** Zoom-in on the InSb/InP interface in (a) (blue square) focusing on the atomic columns using an Inverse Fast Fourier Transform procedure filtered for the (110) periodicity. The vertical lines represent the columns of atoms which, at the interface of InSb/InP, show misfit dislocations, encircled in red. Counting the ratio of InSb columns and InP columns gives a lattice mismatch of 10.34%. **d:** Fast-Fourier Transform of the InSb nanowire (a, green), the InSb/InP interface (a, blue) and the InP substrate (a, purple), demonstrating that the InSb and InP regions are both single crystalline. The FFT of the InSb/InP interface region shows double spots indicating two different lattice parameters. **e**: Cross-section TEM image of an InSANe InSb nanowire grown and imaged along the [0-11] direction. **f**: Zoom in of (**e**) where a pair of horizontal twin planes can be observed a few nanometres above the InP/InSb interface. **g**: Cut along a <112> direction grown wire, looking in the perpendicular <110> zone axis. A few nanometres above the InP/InSb interface, a horizontal twin plane can be observed along the entire observed length of the wire, indicated here by a red arrow.

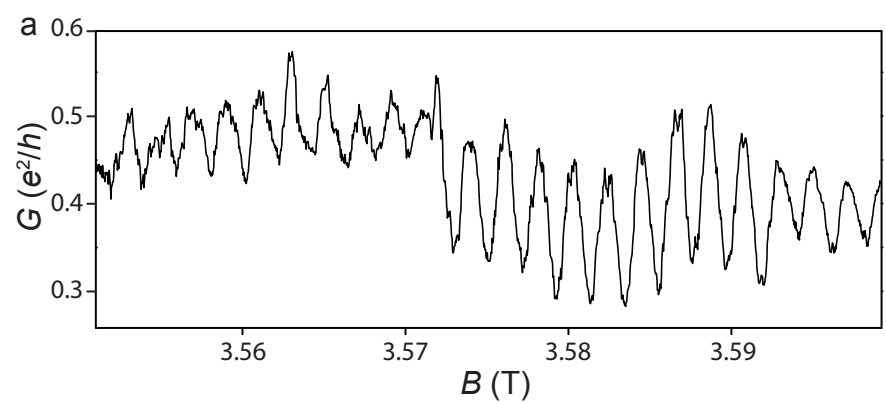
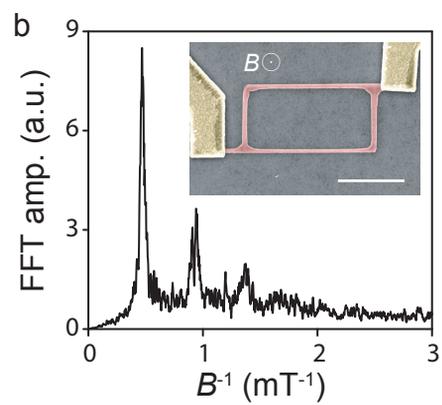
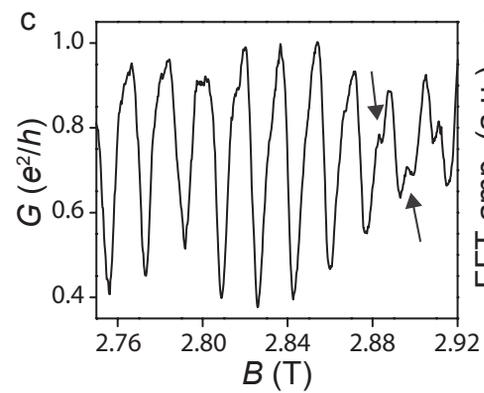
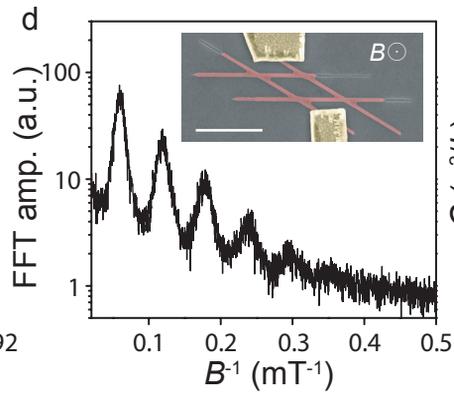
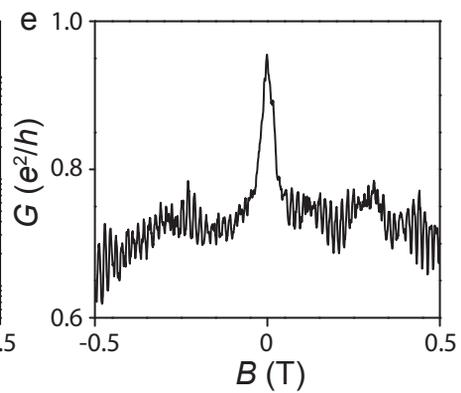

**Figure 4| Aharonov-Bohm and weak-anti-localization effects in InSANe nanowire loops.**

**a**: Magneto-conductance of device A shows Aharonov-Bohm (AB) oscillations with a period of ~ 2mT. **b**: Fast Fourier Transform (FFT) spectrum of the AB oscillations from device A, indicating the frequency peaks up to the 3rd order harmonic. Inset, false-colour SEM image of the device. An InSb nanowire loop (red) is in contact with normal metal electrodes Cr/Au (yellow) with an out-of-plane magnetic field and a fridge temperature of 20mK. The device has a global top gate that is not shown in the SEM image. Scale bar is 1 µm. The measured AB period matches the loop area of ~ 2 µm$^2$. **c**: Magneto-conductance of device B shows a larger AB period and oscillation amplitude (~0.6 $e^2/h$), due to its smaller loop area compared to device A. The arrows indicate oscillations due to higher AB harmonics. **d**: FFT spectrum of device B (SEM in inset with scale bar 1 µm) showing up to five harmonic peaks. **e**: Magneto-conductance of device B (ensemble averaged) shows a sharp weak anti-localization peak at $B$ = 0T.

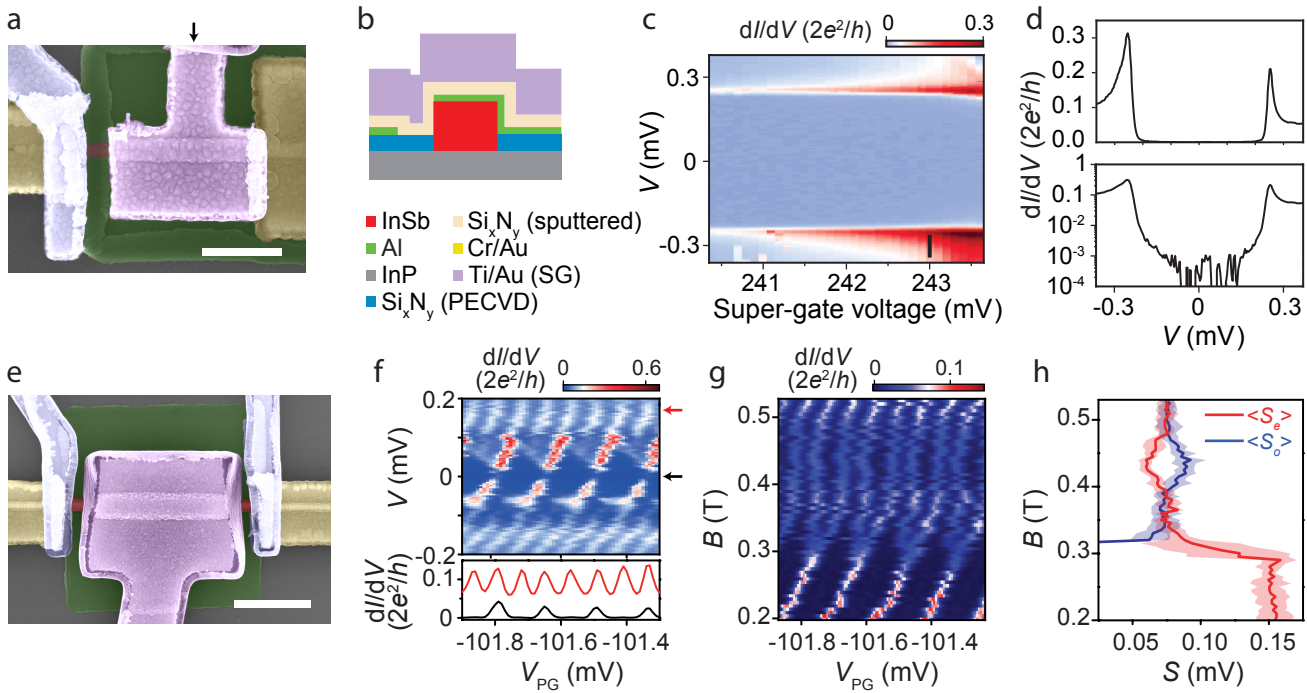

**Figure 5| Hard gap and 2e-periodic Coulomb blockade in InSANe InSb-Al hybrid nanowire devices.**
**a:** False-colour SEM of a typical N-NW-S device, with a schematic of the device cross-section (**b**) at the position indicated by the black arrow in (**a**). The InSb nanowire (red) is covered by a 7 nm thick Al layer (green, covered by the etch mask). Part of the Al film on the nanowire is selectively etched (see methods) before normal metal electrodes deposition (yellow, Cr/Au) and gate-tuneable tunnel barrier. The tunnel- and super-gates are Ti/Au (blue and purple respectively), deposited on top and separated from the wire by a $Si_xN_y$ dielectric layer as seen in the cross-section schematic on the right. Scale bar is 500 nm. Fridge temperature is 20 mK. **c**: Differential conductance (d$I$/d$V$) as a function of bias voltage ($V$) and super-gate voltage in the tunnelling regime, resolving a hard superconducting gap ($\Delta$~250 $\mu$eV) with the line-cuts (both linear and logarithmic scale) shown in **d** at gate voltage indicated by black bar in **c**. The sub-gap/above-gap conductance suppression reaches two orders of magnitude. **e**: False-colour SEM of the superconducting island device. The Al island on the nanowire is around 1 $\mu$m long, with a top plunger gate (purple) to tune the electron density, and two tunnel-gates (blue) to control the tunnel coupling to the two leads. Scale bar is 500 nm. **f:** Differential conductance of the island device as a function of bias and plunger-gate voltage resolving the Coulomb blockade diamonds. The horizontal line-cut at zero bias (black curve) shows 2$e$-periodic Coulomb oscillations where each peak corresponds to adding/removing two electrons (one Cooper pair), suggesting negligible quasi-particle poisoning. At higher bias voltage where quasi-particle can be excited, the Coulomb oscillations become 1$e$-periodic. **g**: Magnetic field dependence of the Coulomb oscillations at zero bias voltage with the field direction along the wire. The 2$e$-peaks split into 1$e$-peaks at around 0.3 Tesla, indicating a sub-gap state crosses zero energy. **h**: even ($S_e$ red) and odd ($S_o$ blue) peak spacing extracted from (**g**), with error bars indicated with shaded areas, showing possible Majorana or Andreev oscillations.

# Supplementary information: In-plane Selective Area InSb-Al Nanowire Quantum Networks

Roy L.M. Op het Veld[1,2†], Di Xu[2†], Vanessa Schaller[2], Marcel A. Verheijen[1,3], Stan M.E. Peters[1], Jason Jung[1], Chuyao Tong[1], Qingzhen Wang[2], Michiel W.A. de Moor[2], Bart Hesselmann[2], Kiefer Vermeulen[2], Jouri D.S. Bommer[2], Joon Sue Lee[4], Andrey Sarikov[5,6], Mihir Pendharkar[9], Anna Marzegalli[7] Sebastian Koelling[1], Leo P. Kouwenhoven[2,8], Leo Miglio[5], Chris J. Palmstrom[4,9,10], Hao Zhang[11,12,2*], Erik P.A.M. Bakkers[1*]

[1] *Department of Applied Physics, Eindhoven University of Technology, 5600MB Eindhoven, The Netherlands*
[2] *QuTech and Kavli Institute of Nanoscience, Delft University of Technology, 2600GA Delft, The Netherlands*
[3] *Eurofins Materials Science Eindhoven, High Tech Campus 11, 5656AE Eindhoven, The Netherlands*
[4] *California NanoSystems Institute, University of California, Santa Barbara, California 93106, USA.*
[5] *L-NESS and Dept. of Materials Science, University of Milano-Bicocca, I-20125 Milano, Italy.*
[6] *V. Lashkarev Institute of Semiconductor Physics, National Academy of Sciences of Ukraine, Kiev, Ukraine.*
[7] *L-NESS and Dept. of Physics, Politecnico di Milano, I-22100 Como, Italy*
[8] *Microsoft Quantum Lab Delft, 2600GA Delft, The Netherlands*
[9] *Electrical and Computer Engineering, University of California, Santa Barbara, California 93106, USA*
[10] *Materials Department, University of California, Santa Barbara, California 93106, USA*
[11] *State Key Laboratory of Low Dimensional Quantum Physics, Department of Physics, Tsinghua University, Beijing 100084, China*
[12] *Beijing Academy of Quantum Information Sciences, Beijing 100193, China*

[†] *These authors contributed equally to this work.*

[*] *Corresponding author e-mail: hzquantum@mail.tsinghua.edu.cn, e.p.a.m.bakkers@tue.nl*

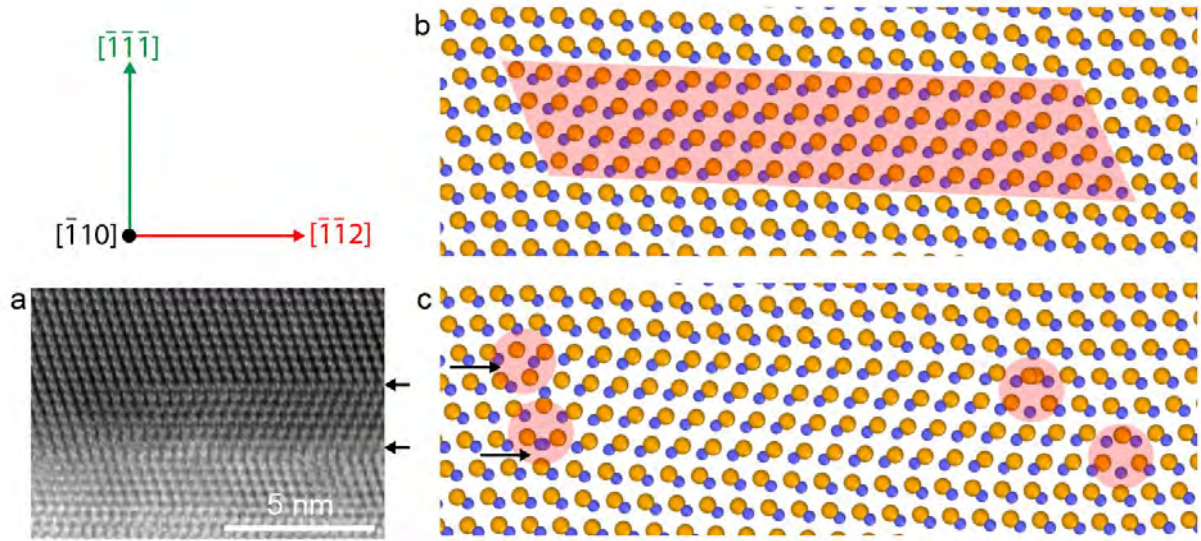

**Figure S1 Molecular Dynamics simulation on crystal relaxation between two twinned regions.**

**a**: cross section TEM image looking along the <110> zone axis, with a twinned region on the right, indicated by the black arrows. The twinned region is not present on the left side of the image. In order to understand how the twins are connected to the original crystal orientation at their lateral boundaries, atomic simulations are performed using the semi-empirical Vashishta energy potentials for InP, which takes the first- and second-nearest neighbours into account. There is no potential available for InSb, but there is no general limitation, since we are looking for crystallographic ordering, which is the same for both semiconductors. **b**: Crystal cell containing a twinned region (marked in red, starting at the black arrows, with the same orientation as the TEM image) with lateral boundaries parallel to the <110> direction prior to molecular dynamics (MD) annealing and energy minimization. **c**: crystal cell after MD annealing and energy minimization. $90^0$ extrinsic stacking fault dislocation lines are formed parallel to the <1-10> direction (highlighted by red circles in the schematic cross-section), each composed by two $30^0$ Shockley partial dislocations located in consecutive lattice planes. This is the only way a (111) oriented twin in the InSb lattice can be terminated and connected to the original crystal orientation, without forming amorphous regions at their boundary. It should be noted that raising the annealing temperature to 1000 K did not change the final structure of the InP cell in **(b)**. In conclusion, twins are always accompanied by dislocation lines running along all the <110> directions in the basic (111) plane, providing additional relieve of the residual misfit stain. This enables the InSb crystal above the twin plane to be completely free of any strain-induced defects.

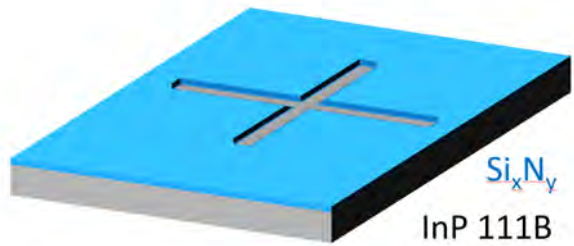

**Figure S2 Substrate after processing**

An InP (111)B undoped 2-inch wafer is used as the growth substrate for InSANe InSb nanowires. First, the native surface oxides are removed by dipping the wafer in $H_2O:H_3PO_4$ 10:1 for 3 minutes. Next, a 20 nm $Si_xN_y$ layer is deposited using plasma enhanced chemical vapor deposition (PeCVD). Resist (ZEP520A +C60) is spun at 2000RPM for 60 sec and then baked at 150 $^0$C for 15 min. Electron beam lithography is used to define the structures. The wafer is developed using n-amyl acetate for 60 sec, followed by an Methyl-iso-butylketon: isopropanol (MIBK:IPA 89:11) mixture for 45 seconds and finally an IPA rinse. Next, the $Si_xN_y$ is locally removed using reactive ion etching with $CHF_3$, an anisotropic dry etching technique. The resist is stripped, and the wafer cleaned by oxygen plasma exposure. A final $H_2O:H_3PO_4$ 10:1 clean is performed for 3 minutes, in order to remove the surface oxide on the InP substrate prior to growth.

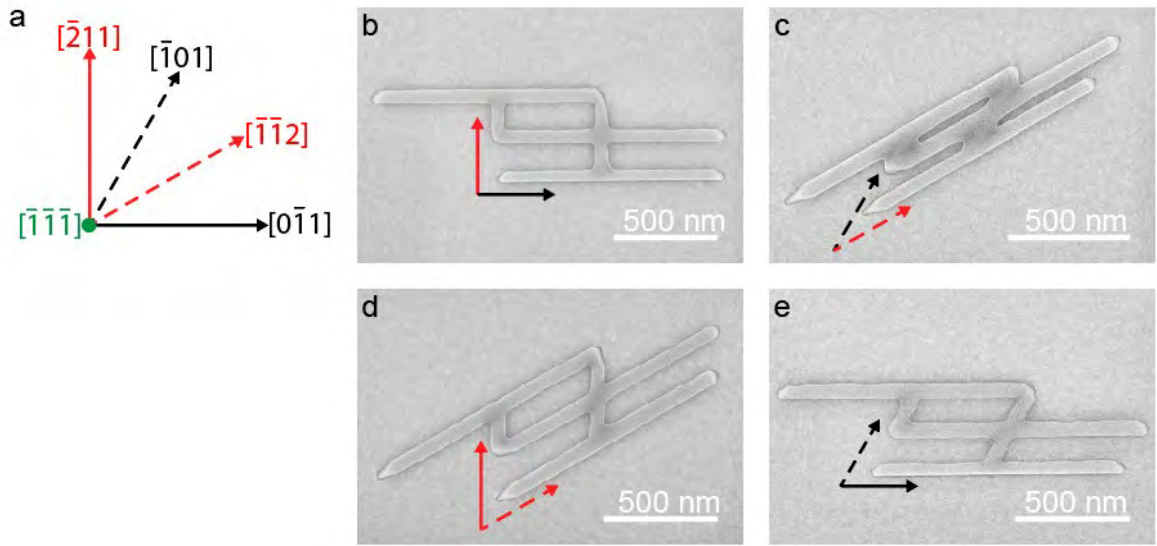

**Figure S3 Sensitivity to crystal directions**

In-plane InSb nanowires can be grown in specific crystal directions, namely the <112> and the <110> families of directions. In order to create in-plane networks, mask openings must be in any combination of these directions. **a**: Crystal directions on a (111)B substrate showing the first quadrant of <112> and <110> directions. Combinations can be made with <112> and <110> directions in either 90 degrees **(b)** or 30 degrees **(c)**. under 60 degrees angles combinations of <112> directions **(d)** or <110> directions **(e)** can be defined.

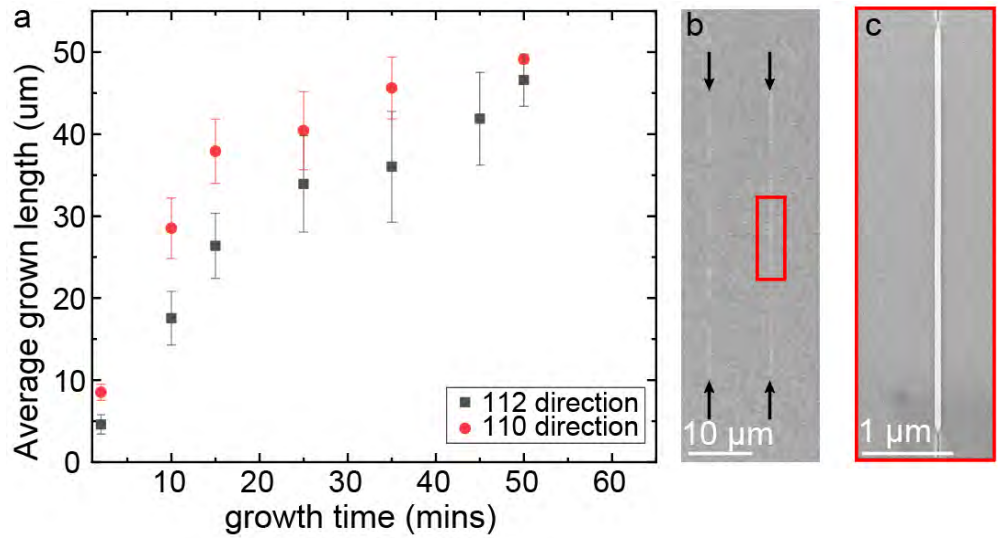

**Figure S4: crystal direction dependent grown length as a function of growth time**

**a**: Average grown length of 50 um long openings in the mask in the <112> (black squares) and <110> (red circles) crystal directions as a function of growth time. In all cases the same temperature of 470 $^0$C and V/III ratio of ~8000 is used during growth in the MOVPE. **b**: SEM image showing two 50 μm lines (indicated with black arrows) in the <112> direction after 35 minutes of growth. A zoom-in is shown in (**c**), clearly showing a grown segment on InSb nanowire (bright white), with at the top and bottom of the image a small segment of line in the mask that is not yet grown.

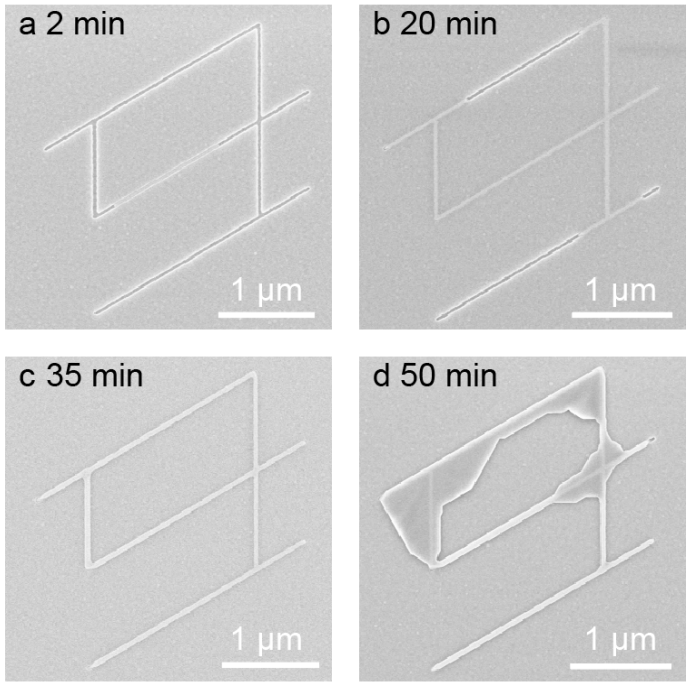

**Figure S5: Network growth as a function of growth time**

**a-d**: SEM image of a representative structure at different growth times. The growth starts with one nucleation site **(a)** and starts growing outwards, following the openings in the mask **(b)**. Once the structure is fully filled, it starts to grow only in the vertical direction **(c)**. When growth is not stopped in time, the structure starts growing out of and over the mask **(d)**.

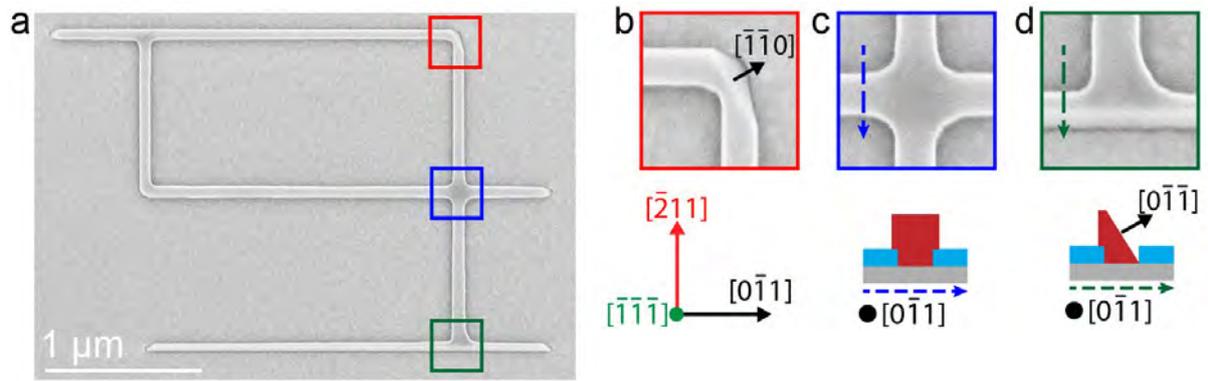

**Figure S6 Influence of extended arms on terminating side facets.**

**a**: Scanning electron microscope image of an InSb nanowire network in a single qubit design structure. The terminating <110> side facets shown in figure 1 are visible in **(b)** (red square) and **(d)** (green square). **b**: The top-right corner of the structure shows the [-1-10] terminating facet. **c**: The cross-junction is missing the [0-1-1] terminating side facet, which is present in **(d)**. The extra arm extending down, changes the preferred terminating side facet here. A sketch is shown for both **(c)** and **(d)** to illustrate the difference in geometry.

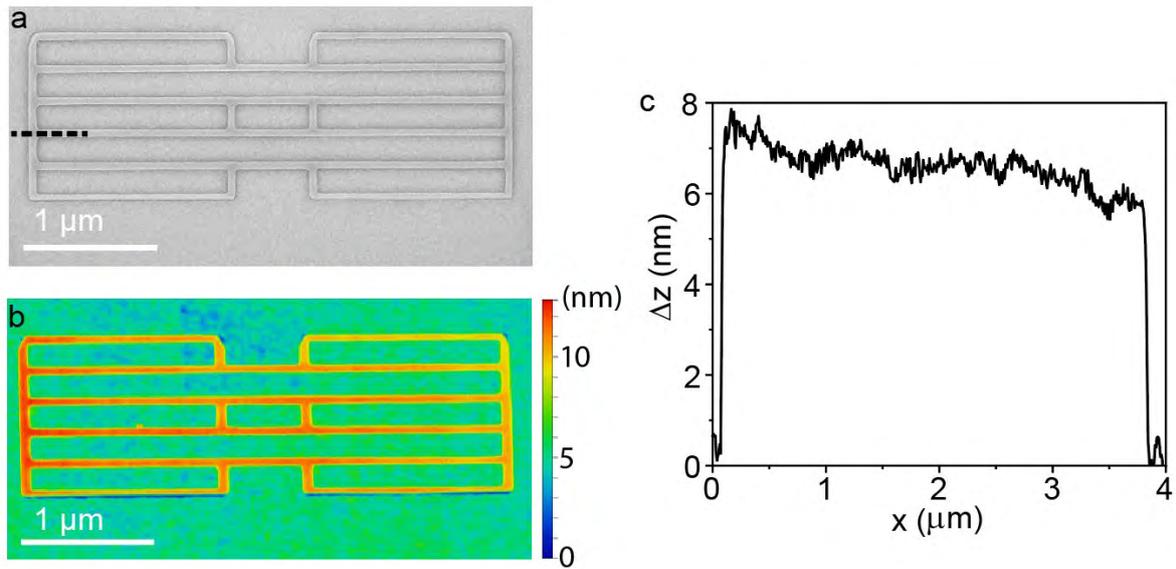

**Figure S7 Atomic force microscopy analysis of in-plane selective-area InSb network**

**a**: SEM image of a 4-qubit design. **b**: AFM height colour map of the same structure as in **(a)** showing the uniform height of the InSb structure. **c**: Line cut taken through the 4th line in the [110] direction showing the height relative to the top surface of the $Si_xN_y$ mask. The height changes ~1.5 nm over a length of 4 µm, or 1 atomic step per 2 µm indicating a layer by layer growth on the top (111)B surface of the structure.

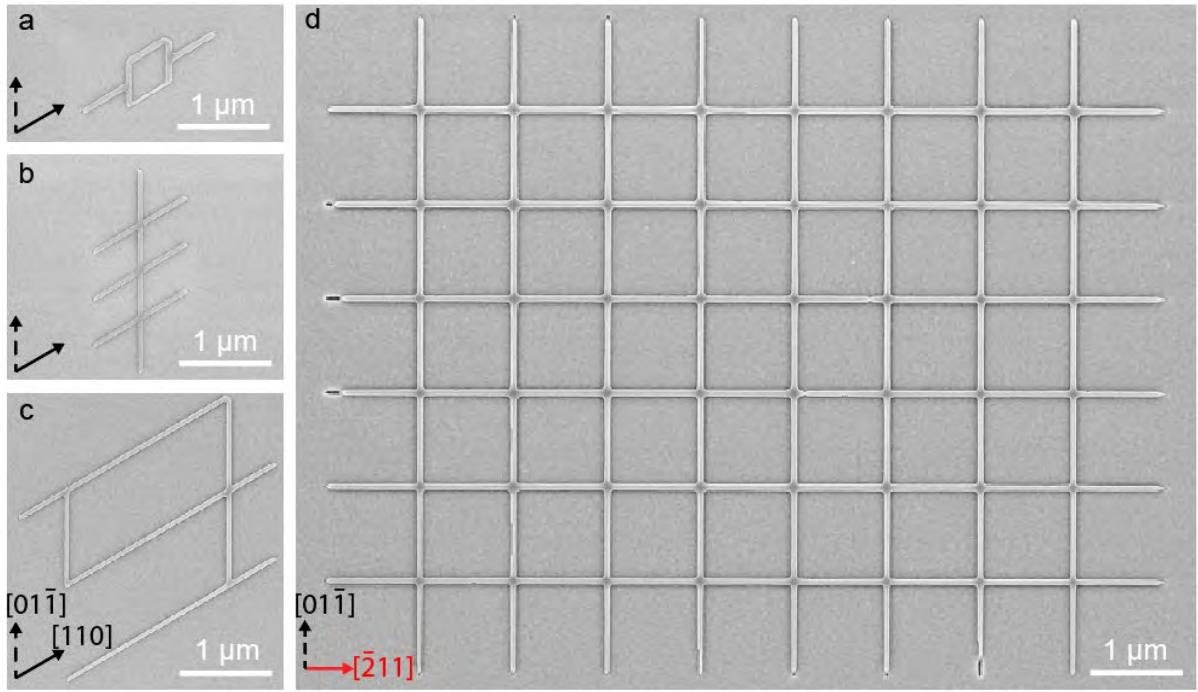

**Figure S8 Scalability of In-plane Selective Area Networks of InSb nanowires**

To demonstrate the scalability of this growth technique, different network designs in different sizes have been grown on the same substrate in the same growth run. All scalebars are the same size and represent 1um, showing that the growth of 200 nm by 200 nm loops **(a)**, up to 11um by 11um hashtag networks **(d)** can be grown at the same time, independent of crystal direction, given that the mask is opened in <112> and/or <110> crystal directions.

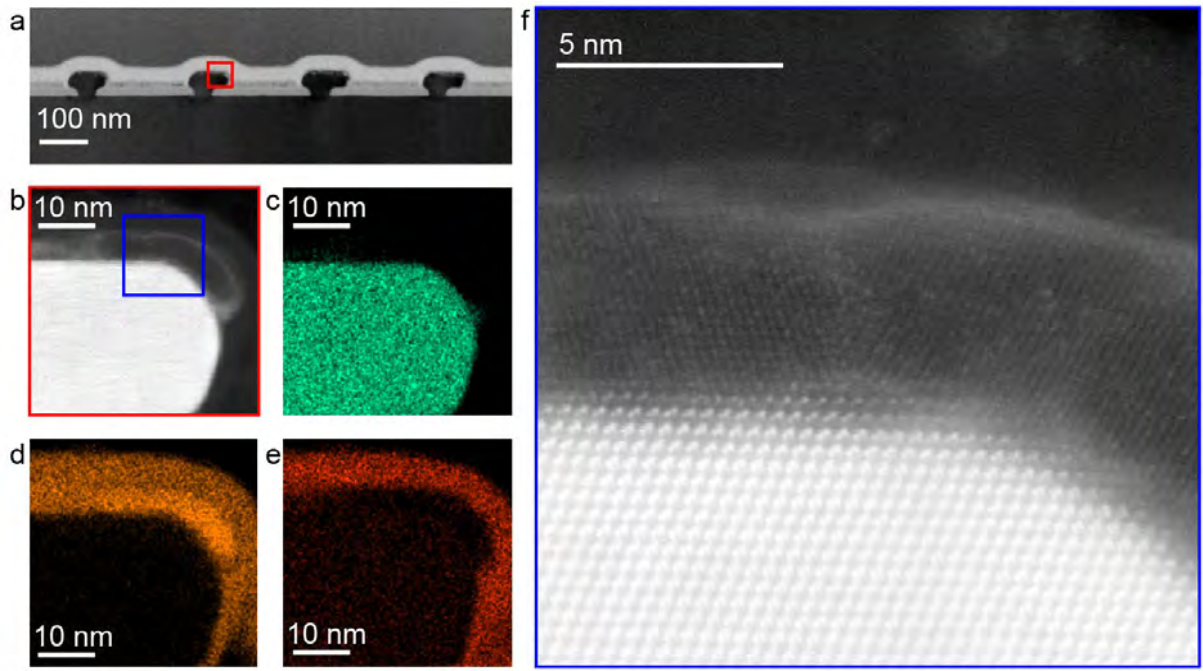

**Figure S9 Al interface in the InSANe InSb-Al hybrid system**

**a**: cross-section TEM view of four InSANe InSb nanowires grown on top of an InP (111)B substrate. **b**: Zoom-in of the red area in **(a)**, showing the InSb nanowire top edge with a 5 nm Al layer deposited on the wire. EDX analysis of the area in **(b)** shows the Sb (**c**, green), Al (**d**, orange) and O (**e**, red) in the InSb-Al hybrid network, depicting an oxide free InSb-Al interface. **f**: A zoom-in on the interface in **(b)** shows the clean and smooth interface between the InSb wire and the Al. The crystal lattice of two separate Al grains can be identified in the image, showing the high quality of the Al layer.

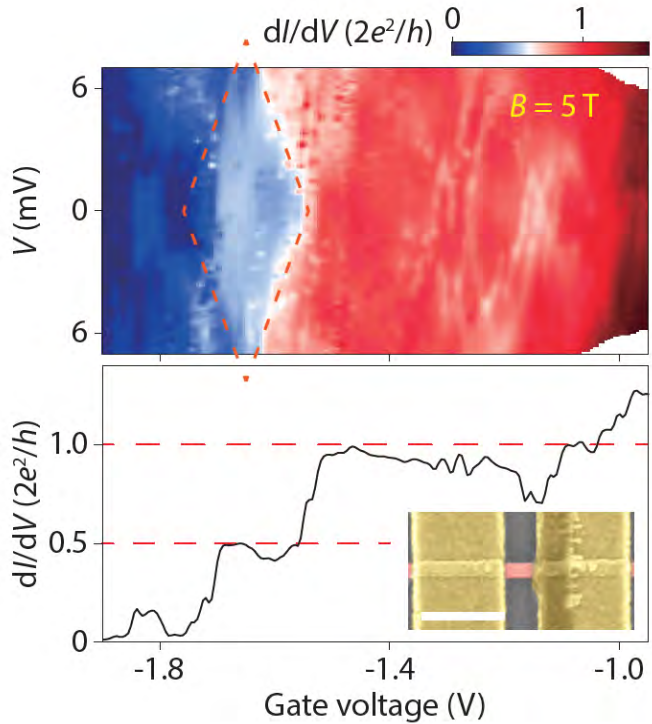

**Figure S10 Quantized conductance of an in-plane InSb nanowire.**

Upper panel: differential conductance measured as a function of bias voltage *V* and gate voltage at a magnetic field at 5T. Conductance plateaus are visible, and the plateau of $e^2/h$ is indicated with red dashed lines. A g-factor of around 57 can be roughly extracted. Lower panel: horizontal line cut of the upper panel at *V*=0. Conductance at $e^2/h$ and $2e^2/h$ is marked with red dashed lines. We note the plateau quality is expected to be worse than VLS InSb nanowires,[1] probably due to the facts that 1) the nanowire was heavily etched by Ar plasma before contact deposition; 2) near the pinch-off region, the top gate pushes the electron wavefunction towards the nanowire bottom (substrate with large lattice mismatch). Both introduce extra disorders which degrade the plateau quality. Inset: SEM image of the quantum point contact device. A nanowire segment (red) is contacted by Au (yellow) with 200 nm spacing. Global dielectric and top gate are not shown in this image. Scale bar is 500nm.

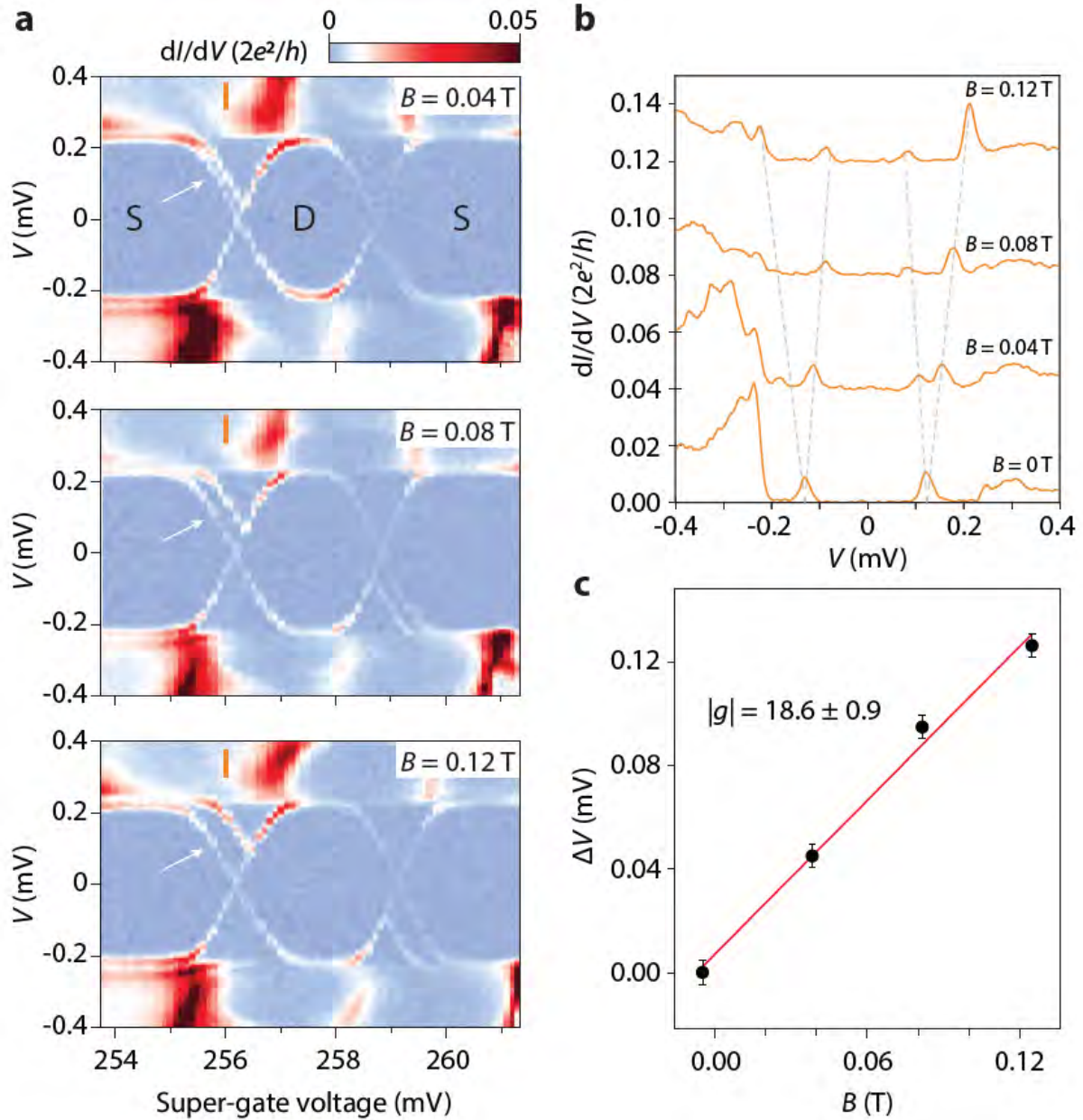

**Figure S11 Extraction of the g-factor in the N-NW-S device.**

a: Differential conductance of the N-NW-S device measured as a function of bias voltage V and super-gate voltage at different magnetic fields (0.04 T, 0.08 T and 0.12 T from top to bottom) along the nanowire. Andreev levels corresponding to the transitions between the singlet and doublet states (indicated with S/D respectively) are observed. [2] Note that the Andreev levels corresponding to singlet ground states have Zeeman splitting (white arrows). **b**: Line traces from panel **(a)** at the super-gate voltage of 256 mV (at orange lines in a), plus the line trace at the same super-gate voltage at zero magnetic field. Offset between lines is $0.04 \times 2e^2/h$. Zeeman splitting is indicated by the dashed lines. **c**: The spacing between the split peaks (blue dots) of **d** (positive side) as a function of B. The slope of the linear fit (green line) is used to extract the g-factor.

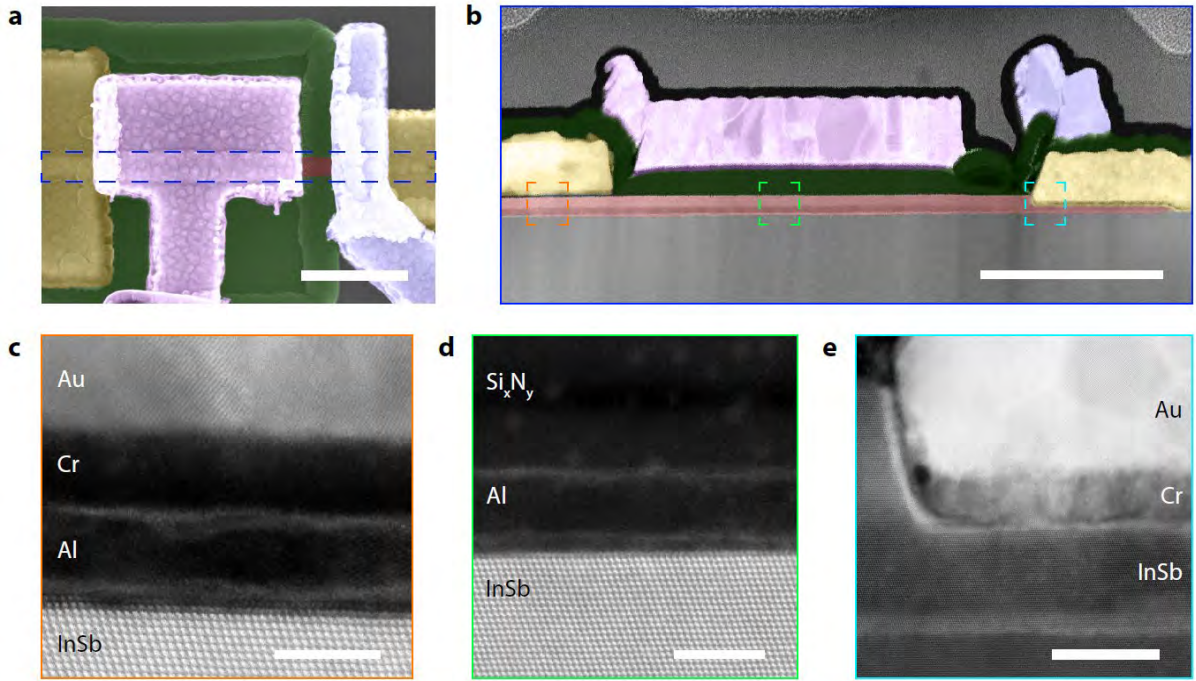

**Figure S12 Cross-section TEM of a N-NW-S device.**

**a**: SEM image of a N-NW-S device. The InSb nanowire (red) is covered by an Al layer (green, covered by the etch mask). Contacts are Cr/Au (yellow). The tunnel- (blue) and super-gates are Ti/Au, deposited on top and separated from the wire by a $Si_xN_y$ dielectric layer. Scale bar is 500nm. The area indicated with blue dashed lines shows where the TEM specimen is made. **b**: Cross-section TEM image of the device. The structures have the same colours as those of the corresponding structures in **(a)**, except for the $Si_xN_y$ dielectric coloured in green together with Al. Scale bar is 500nm. **c**: High resolution TEM image of the orange region in **(b)**, showing the structure of the super contact. Note that the Ti/Au can contact the nanowire from the facet not covered by Al even though the $AlO_x$ layer on top of Al is not sufficiently removed by a milling process. **d**: High resolution STEM image of the green region in **(b)**, showing the structure of the InSb-Al-$Si_xN_y$ dielectric stack. **e**: High resolution STEM image of the cyan region in **(b)**, showing the structure of the normal contact. Note that the milling process before depositing Cr/Au has etched into the InSb sufficiently to form an oxygen-free contact interface.

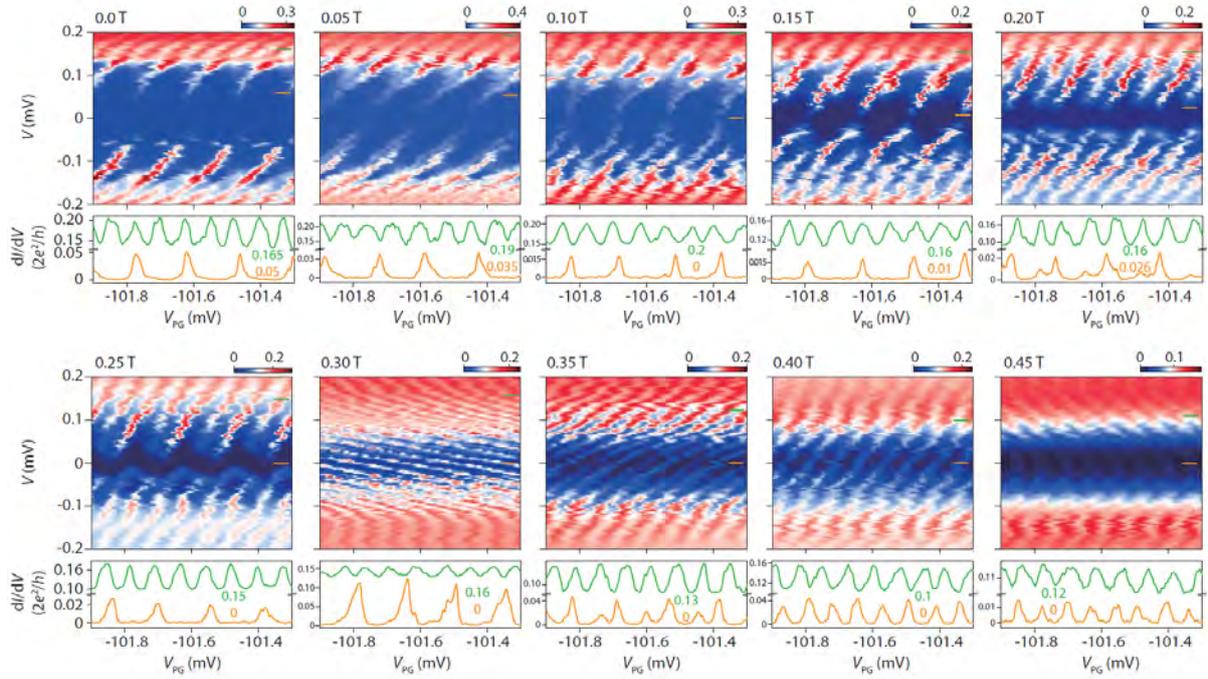

**Figure S13 Coulomb diamond of InSb-Al island device: transition from 2*e*-periodic to 1*e*-periodic.**

Differential conductance as a function of plunger gate $V_{PG}$ and bias voltage $V$ at different fixed parallel magnetic fields (from 0 T to 0.45 T, with 0.05 T step). The Coulomb diamond is not symmetric due to the asymmetric tunnelling strength of the two barriers. The panels below show line cuts at low bias (orange lines) and high bias (green lines) with the bias voltage indicated (in units of mV). The transition from 2*e*-periodic to 1*e*-periodic oscillations can be observed at low bias voltages.